\documentclass[journal,article,submit,moreauthors,pdftex,12pt,a4paper]{mdpi} % for use with pdfLaTeX only
\lastpage{x}
\doinum{10.3390/------}
\pubvolume{xx}
\pubyear{2012}
\history{Received: xx / Accepted: xx / Published: xx}

\usepackage{amssymb,amsmath}
\usepackage{graphicx}
\usepackage{dsfont}
\newcommand{\re}{\mbox{$\rm e$}}
\newcommand{\ri}{\mbox{$\rm i$}}
\newcommand{\rd}{\mbox{$\rm d$}}
\Title{Information Geometry of Complex Hamiltonians and Exceptional Points}

\Author{Dorje C. Brody $^{1,\star}$ and Eva-Maria Graefe $^{2}$}

\address{%
$^{1}$ Mathematical Sciences, Brunel University, Uxbridge UB8 3PH, UK \\
$^{2}$ Department of Mathematics, Imperial College London, London SW7 2AZ, UK}

\corres{Dorje.Brody@brunel.ac.uk}%, Tel. and Fax number of the corresponding author.}

\abstract{
Information geometry provides a tool to systematically investigate parameter sensitivity of 
the state of a system. If a physical system is described by a linear combination of eigenstates 
of a complex (that is, non-Hermitian) Hamiltonian, then there can be phase transitions where 
dynamical properties of the system change abruptly. In the vicinities of the transition points, the 
state of the system becomes highly sensitive to the changes of the parameters in the Hamiltonian. 
The parameter sensitivity can then be measured in terms of the Fisher-Rao metric and the 
associated curvature of the parameter-space manifold. A general scheme for the geometric study 
of parameter-space manifolds of eigenstates of complex Hamiltonians is outlined here, leading to 
generic expressions for the metric. 
}


\begin{document}
%%%%%%%%%%%%%%%%%%%%%%%%%%%%%%%%%%%%%%%%%%%%%%%%%%%%%%%%%%%%

\section{Introduction}

In statistical physics, if a system is in equilibrium with a heat bath at inverse temperature $\beta$, 
then the state of the system is characterised by the canonical phase-space density function 
\begin{eqnarray}
\rho(x|\beta) = \frac{\re^{-\beta H(x)}}{Z(\beta)}, 
\label{eq:1}
\end{eqnarray}
where $H(x)$ is the Hamiltonian function on phase space ${\mathit\Omega}$ and the partition 
function $Z(\beta)$ is given by the integral of the Boltzmann weight $\exp(-\beta H)$ over 
${\mathit\Omega}$. Systems having sufficiently rich inter-particle interactions can exhibit phase 
transitions. Typically a phase 
transition is associated with the breakdown of the analyticity of one or more thermodynamic 
quantities, such as specific heat or magnetic susceptibility. The density function (\ref{eq:1}), on 
the other hand, is analytic, and one can legitimately ask in which way a breakdown of analyticity 
can be extracted from an analytic quantity. Indeed, for a real analytic function it is not possible to 
find a breakdown of analyticity in a system having finitely many degrees of freedom, and it is 
mandatory to consider the operation of a thermodynamic limit. 

If the state of a system is described by a density function (or a discrete set of probabilities) that 
lacks analyticity in the first place, however, then a phase transition can be seen without involving 
the mathematically cumbersome operation of thermodynamic limit. Such situations arise in many 
physical contexts. For example, if an isolated quantum system is in a microcanonical state having 
support on the level surface of the expectation of the Hamiltonian, then the density of states is not 
analytic and one can find thermal phase transitions in small quantum systems \cite{DCB2}. 

Another important example arises when considering eigenstates, or linear combinations of them, of complex Hamiltonian operators in quantum mechanics. While quantum mechanics traditionally focusses on closed systems described by Hermitian Hamiltonians, recently there has been considerable interest in relaxing the Hermiticity condition to consider more general complex Hamiltonians. For a Hermitian Hamiltonian in finite dimensions, the associated 
eigenfunctions are analytic in the parameters of the Hamiltonian, and a breakdown of analyticity 
may be obtained only in infinite dimensions. In the case of a complex Hamiltonian, however, the 
associated eigenfunctions need not be analytic in the parameters of the Hamiltonian, and phase 
transitions can be seen in finite matrix Hamiltonians. This situation is reminiscent of the analysis 
proposed by Lee and Yang \cite{YL,LY} where the breakdown of analyticity associated with the 
canonical density function (\ref{eq:1}) can be explained by extending the parameters into a complex 
domain (see, e.g., \cite{BE} for a heuristic but informative exposition of the Lee-Yang theory). In 
this case, the canonical density function can exhibit lack of analyticity even in a system with 
finitely many degrees of freedom, in a way that resembles the eigenstates of finite complex 
Hamiltonians (see also \cite{CHM} for a related point of view on these issues). 

The transition points, or critical points, associated with eigenstates of a complex Hamiltonian 
are points at which degeneracies occur, that is, points at which not only eigenvalues but also 
eigenstates coalesce. In the literature, these critical points are often referred to as `exceptional 
points' (see \cite{Heiss} for a concise and informative overview of the physics of exceptional 
points). The purpose of the present paper is to investigate properties of eigenstates of complex 
Hamiltonians around exceptional points, from the viewpoint of inference theory. If a system, say, 
is in an eigenstate of a complex Hamiltonian, 
but an experimentalist does not know the exact values of the parameters in the Hamiltonian, then 
these parameter values can be estimated from observational data. Inference theory concerns the 
analysis of this data and in particular error bounds associated with such estimates. Evidently, if 
the eigenstate is sensitive to the changes of the parameter values, that is, if the state of the 
system changes significantly when the Hamiltonian is modified only slightly, then in this regime it 
is easy to estimate the parameter values. Conversely, if the state of the system is almost 
unaltered under the changes of the parameters, then in this regime estimation errors will be large. 
Hence from the viewpoint of inference theory we are interested in identifying parameter sensitivity 
of the eigenfunctions of complex Hamiltonians. The method of information geometry then allows 
us to proceed with such an analysis, since it assigns distance measures between eigenstates 
of the Hamiltonian associated with different parameter values. 

The paper is organised as follows. In \S\ref{sec:2} we give a brief overview of the method of 
information geometry applied to statistical physics for the benefit of readers less acquainted with 
the material. For a recent review of the use of information geometry in statistical mechanics, and 
a comprehensive list of references, see \cite{BH}. In \S\ref{sec:3} we explain in which way the 
standard method of information geometry, based on structures of \textit{real} Hilbert space, extends 
into the case of a \textit{complex} Hilbert space. As an example we consider in \S\ref{sec:4} the 
Hilbertian manifold generated by an eigenstate of a Hermitian Hamiltonian. In this context we make 
use of the idea proposed in \cite{ZGC} of applying first-order perturbation theory to deduce a 
generic and intuitive form of the Riemannian metric on the manifold. Furthermore, we derive a new 
type of quantum uncertainty relations that arises naturally from the estimation of parameters in 
the Hamiltonian. In \S\ref{sec:5} we turn to the analysis of the geometry of the manifold associated 
with eigenstates of a complex Hamiltonian. Specifically, we derive an expression for the metric 
using the Rayleigh-Schr\"odinger perturbation theory on complex Hamiltonians, away from 
exceptional points. This perturbation expansion, however, breaks down in the vicinity of 
exceptional points. Thus in \S\ref{sec:6} we apply generalised perturbation theory so as to 
identify the generic structure of the metric close to an exceptional point where two of the 
eigenstates coalesce. In \S\ref{sec:7} we work out nonperturbatively the metric geometry of the 
parametric eigenstates in a simple example system, showing the existence of a geometric 
singularity at the exceptional point. The result is also compared to the perturbative analysis of 
\S\ref{sec:6}. 

We remark that while physical effects associated with the existence of exceptional points have been observed in laboratory experiments already as early as 1955 \cite{Panc55}, only relatively recently a controlled experimental investigation is being pursued \cite{Richter,Richter2,Lee09}. In particular, 
investigations into properties of complex Hamiltonians have increased significantly over the past 
decade since the observation of Bender and Boettcher that complex Hamiltonians possessing 
parity-time (PT) reversal symmetry can possess entirely real eigenvalues \cite{BB}. Phase 
transitions associated with the breakdown of PT symmetry of the eigenfunctions at exceptional points have also been predicted or observed in a range of model systems and experiments 
\cite{DC1,NM,AM,DC2,DC3,SR,TK,SR2,TK2,BD,BG2,CMB}, and constitute an interesting and 
exciting area of application of information geometry. It is our hope that the 
present paper serves as a  concise introduction to the physics of complex Hamiltonians for those 
who work in the area of information geometry, and at the same time an introduction to information 
geometry for those who work in the study of physical systems described by complex Hamiltonians.

\section{Information geometry and statistical mechanics}
\label{sec:2}

To gain visual insights into the nature of critical points in statistical physics we follow the mathematical 
scheme proposed by Rao \cite{Rao} and consider the square-root map 
\begin{eqnarray}
\rho(x|\beta) \to \xi(x|\beta) = \sqrt{\rho(x|\beta)} .
\label{eq:2} 
\label{eqnarray}
\end{eqnarray}
It should be evident that the function $\xi(x|\beta)$ belongs to a real Hilbert space ${\mathcal H}$ of 
square-integrable functions on the phase space ${\mathit\Omega}$ of a given system. Thus, with 
respect to any given choice of coordinates in ${\mathcal H}$ we can think of $\xi(x|\beta)$ for each 
fixed value of $\beta$ as a vector $|\xi(\beta)\rangle\in{\mathcal H}$ of unit length satisfying 
$\langle\xi(\beta)|\xi(\beta)\rangle=1$. Here we use the Dirac notation for 
representing elements of ${\mathcal H}$. If we vary the inverse temperature $\beta$, then the vector 
$|\xi(\beta)\rangle$, representing the thermal equilibrium state at $\beta$, traverses along a smooth 
curve on the unit sphere in ${\mathcal H}$. In particular, in the limit $\beta\to\infty$ the equilibrium 
state $|\xi(\beta)\rangle$ of the system approaches the `ground state' of the Hamiltonian, i.e. a state 
with minimum energy. (Note that unless the parameter is changed adiabatically, the physical state of 
the system will not traverse the path $|\xi(\beta)\rangle$ since a rapid 
change of temperature momentarily brings the state of the system out of equilibrium. Hence we are 
not concerned here with the out-of-equilibrium dynamics of the system as such. Rather, we are 
interested in how an equilibrium configuration at one temperature is related to the equilibrium 
configuration at another temperature, and this is characterised by the path $|\xi(\beta)\rangle$. 
Our analysis thus reproduces the dynamical theory only in the adiabatic limit.)

If a system exhibits phase transitions, then the equilibrium curve $|\xi(\beta)\rangle$ can branch 
out into several curves at the critical points. For example, in the case of the Ising 
model with vanishing magnetic field the curve bifurcates at the critical temperature; for the van der 
Waals model the curve trifurcates at the critical point. Of course, such a scenario can prevail in the 
context of canonical state (\ref{eq:1}) only if the dimensionality of ${\mathcal H}$ is infinite; 
nevertheless the concept of a one-dimensional curve residing on an infinite-dimensional unit sphere 
offers a visual characterisation of the situation. 

In the case of a one-parameter family of states $|\xi(\beta)\rangle$ the parametric sensitivity can 
be measured in terms of the squared `velocity' (metric) and the squared `acceleration' (curvature) 
of the curve. By squared velocity, which we shall denote by $G$, we mean the inner product 
\begin{eqnarray}
G = 4 \langle{\dot\xi}(\beta)|{\dot\xi}(\beta)\rangle, 
\label{eq:3}
\end{eqnarray}
where the dot represents differentiation with respect to $\beta$, and the factor of four is purely 
conventional so that $G$ agrees with the information measure introduced by Fisher \cite{fisher}. 
A short calculation shows \cite{BH0} that in the case of the canonical state (\ref{eq:1}) we have 
$G=\Delta H^2$, that is, the variance of the Hamiltonian in the canonical state (\ref{eq:1}). 
Hence in a region where the equilibrium energy uncertainty is small, the state of the system does 
not change much when the inverse temperature is changed, and this in turn means that an accurate 
estimation of $\beta$ is difficult. Indeed, from the Cram\'er-Rao inequality one finds that the 
quadratic error $\Delta\beta^2$ of the estimation is bounded below by $(4\Delta H^2)^{-1}$, 
and one obtains the thermodynamic uncertainty relation \cite{Mandelbrot,BH3}: 
\begin{eqnarray}
\Delta T^{-1} \, \Delta H \geq \frac{k_B}{2} , 
\end{eqnarray}
where we have written $\beta=1/k_B T$, with $k_B$ the Boltzmann constant. 

Similarly, the acceleration vector $|\alpha(\beta)\rangle$ of the curve is defined by 
\begin{eqnarray}
|\alpha(\beta)\rangle = |{\ddot\xi}(\beta)\rangle - 
\frac{\langle{\dot\xi}(\beta)|{\ddot\xi}(\beta)\rangle}{\langle{\dot\xi}(\beta)|{\dot\xi}(\beta)\rangle} 
\, |{\dot\xi}(\beta)\rangle - \langle\xi(\beta)|{\ddot\xi}(\beta)\rangle\, |\xi(\beta)\rangle , 
\label{eq:4}
\end{eqnarray}
where $|{\ddot\xi}(\beta)\rangle=\partial_\beta^2|\xi(\beta)\rangle$. In terms of the acceleration 
vector the intrinsic curvature ${\mathcal K}^2$ of the curve $|\xi(\beta)\rangle$ is given by
\begin{eqnarray}
{\mathcal K}^2 = \frac{16}{G^2} \langle\alpha(\beta)|\alpha(\beta)\rangle.  
\label{eq:5}
\end{eqnarray}
In the case of the canonical state (\ref{eq:1}) a calculation shows \cite{BH0} that 
\begin{eqnarray}
{\mathcal K}^2 = \frac{\Delta H^4}{(\Delta H^2)^2} - 
\frac{(\Delta H^3)^2}{(\Delta H^2)^3}-1, 
\label{eq:6}
\end{eqnarray}
where we have written $\Delta H^k$ to mean the $k^{\rm th}$ central moment of the 
Hamiltonian in the thermal state (\ref{eq:1}). 
 
More generally, consider a generic density function $\rho(x|\theta)$ dependent on one or several 
parameters $\{\theta^a\}_{a=1,\ldots,N}$, normalised for all values of $\{\theta^a\}$. Then for 
each fixed set of values of $\{\theta^a\}$ the 
square-root map (\ref{eq:2}) determines a point $|\xi(\theta)\rangle$ on the unit sphere 
of Hilbert space. When the values of the parameters are varied, $|\xi(\theta)\rangle$ traverses 
along an $N$-dimensional surface ${\mathfrak M}$ on the sphere. A standard result in 
Riemannian geometry of subspaces then shows that the metric on the subspace ${\mathfrak M}$ 
is determined by 
\begin{eqnarray}
G_{ab} = 4 \langle \partial_a\xi(\theta)|\partial_b\xi(\theta)\rangle, 
\label{eq:7}
\end{eqnarray}
where again the scale factor of four is purely conventional and we have written $\partial_a=
\partial/\partial\theta^a$. The quadratic form (\ref{eq:7}) in the statistical context is known as 
the Fisher-Rao metric. With the expression of the metric tensor (\ref{eq:7}) at hand one 
can proceed to calculate invariant quantities such as the Ricci curvature, or geodesic 
curves on ${\mathfrak M}$. For example, given a pair of points $\theta$ and $\theta'$ on 
${\mathfrak M}$ the separation between the two states $|\xi(\theta)\rangle$ and 
$|\xi(\theta')\rangle$ is given by the distance of the geodesic curve joining these two points 
on ${\mathfrak M}$. Such a distance then determines the divergence measure between the 
two states $|\xi(\theta)\rangle$ and $|\xi(\theta')\rangle$, which is more informative than 
the mere overlap distance $\cos^{-1}(\langle\xi(\theta)|\xi(\theta')\rangle)$. 

In the context of statistical mechanics the curvature of ${\mathfrak M}$ associated with the 
Fisher-Rao metric is singular along the spinodal curve, which contains the critical point \cite{BH}. 
Typically on the equilibrium state manifold ${\mathfrak M}$ there is an unphysical region, e.g., 
a region in which the magnetisation decreases in increasing external field in the Ising model, or 
a region in which, according to equation of states, the pressure decreases in increasing volume 
in the van der Waals model. The spinodal curve gives the boundary of such unphysical regions, 
and it is along this boundary that the curvature diverges, thus in some sense `prevents' a smooth 
entry into unphysical regions. The method of information geometry therefore provides geometric 
insights into the physics of critical phenomena.

\section{Statistical geometry in complex vector spaces}
\label{sec:3}

The geometric analysis of the parametric subspace of the real Hilbert space extends, 
\textit{mutatis mutandis}, to the complex domain---for example, to the complex Hilbert 
space of states in quantum mechanics. There are, however, some modifications arising, 
which we shall discuss now. Consider first the case of a parametric 
curve $|\xi(\theta)\rangle$ satisfying the normalisation condition 
$\langle\xi(\theta)|\xi(\theta)\rangle=1$, where $\langle\xi(\theta)|$ now denotes the Hermitian 
conjugate of $|\xi(\theta)\rangle$. In the complex case the condition $\partial_\theta 
\langle\xi(\theta)|\xi(\theta)\rangle=0$ does not imply $\langle{\dot\xi}(\theta)|\xi(\theta)\rangle=0$ 
owing to the phase factor, so we require a modified expression  
\begin{eqnarray}
|v(\theta)\rangle = |{\dot\xi}(\theta)\rangle - 
\langle{\xi}(\theta)|{\dot\xi}(\theta)\rangle\,  |\xi(\theta)\rangle
\label{eq:8}
\end{eqnarray}
for the proper `velocity' vector. The squared velocity (with a factor of four) is then given by 
\begin{eqnarray}
G = 4\left( \langle{\dot\xi}(\theta)|{\dot\xi}(\theta)\rangle - 
\langle{\xi}(\theta)|{\dot\xi}(\theta)\rangle\langle{\dot\xi}(\theta)|\xi(\theta)\rangle
\right) .
\label{eq:9}
\end{eqnarray}
The simplest situation of a curve $|\xi(\theta)\rangle$ that arises in quantum mechanics is the 
solution to the Schr\"odinger equation 
\begin{eqnarray}
\ri \hbar |{\dot\xi}\rangle = {\hat H}|\xi\rangle 
\label{eq:10}
\end{eqnarray}
with initial condition $|\xi(0)\rangle$ satisfying $\langle\xi(0)|\xi(0)\rangle=1$, where the parameter 
$\theta$ represents time. A short calculation then shows that the squared velocity is given by the 
energy uncertainty: 
\begin{eqnarray}
G = \frac{4 \Delta H^2}{\hbar^2} , 
\label{eq:11}
\end{eqnarray}
which of course is merely the statement of the Anandan-Aharanov relation \cite{AA}. From the 
viewpoint of inference theory we can think of a situation in which a quantum system, prepared in an 
initial state, is made to evolve under the influence of the Hamiltonian ${\hat H}$. After a passage of 
time an experimentalist performs a measurement in order to estimate how much time has elapsed 
since its initial preparation. The Cram\'er-Rao relation then asserts that the quadratic error of time 
estimation is bounded below by $\hbar^{2}(4\Delta H^2)^{-1}$, which is just the Heisenberg 
uncertainty relation (see \cite{holevo,BH0} for further details on the problem of time estimation). 

More generally, an alternative way of deducing the geometry of a parametric subspace 
${\mathfrak M}$ of the quantum state space is to make use of the Fubini-Study geometry of the 
ambient state space. Here, by a `quantum state space' we mean the space of rays through the 
origin of the Hilbert space, i.e. the complex projective space. If we write $\rd s$ for the line 
element on the state space of a neighbouring pair of states $|\xi\rangle$ and $|\xi+\rd\xi\rangle=
|\xi\rangle+|\rd\xi\rangle$, then we have the relation
\begin{eqnarray}
\cos^2 {\textstyle\frac{1}{2}}\rd s = \frac{\langle\xi|\xi+\rd\xi\rangle\langle\xi+\rd\xi|\xi\rangle}
{\langle\xi|\xi\rangle\langle\xi+\rd\xi|\xi+\rd\xi\rangle} . 
\label{eq:14}
\end{eqnarray}
Solving this for $\rd s$ and retaining terms of quadratic order, we obtain the Fubini-Study line 
element
\begin{eqnarray}
\rd s^2 = 4 \frac{\langle\xi|\xi\rangle\langle\rd\xi|\rd\xi\rangle - \langle\xi|\rd\xi\rangle 
\langle\rd\xi|\xi\rangle}{\langle\xi|\xi\rangle^2} . 
\label{eq:15}
\end{eqnarray}
Now suppose that the state $|\xi\rangle=|\xi(\theta)\rangle$ depends smoothly on a set of 
parameters $\{\theta^a\}_{a=1,\ldots,N}$, and is normalised to unity for all values of 
$\{\theta^a\}$. Then we have $|\rd\xi\rangle=|\partial_a\xi\rangle\rd\theta^a$, using the 
summation convention, so that the quantum Fisher-Rao metric on the parameter manifold 
${\mathfrak M}$ induced by the ambient Fubini-Study geometry (\ref{eq:15}) is determined 
by the line element  
\begin{eqnarray}
\rd s^2 = 4\big( \langle{\partial_{a}\xi}|{\partial_{b}\xi}\rangle - 
\langle{\xi}|{\partial_{a}\xi}\rangle\langle{\partial_{b}\xi}|\xi\rangle \big) \rd\theta^a \rd\theta^b .
\label{eq:16_0}
\end{eqnarray}
In other words, the metric tensor is given by 
\begin{eqnarray}
G_{ab} = 4\left( \langle{\partial_{(a}\xi}|{\partial_{b)}\xi}\rangle - 
\langle{\xi}|{\partial_{(a}\xi}\rangle\langle{\partial_{b)}\xi}|\xi\rangle \right) ,
\label{eq:16}
\end{eqnarray}
where the brackets in the subscripts denote symmetrisation (which is just the real part of the 
expression without the symmetrisation). In particular, for $N=1$ we recover the expression in 
(\ref{eq:9}).

\section{Eigengeometry of Hermitian Hamiltonians}
\label{sec:4}

Apart from the examples of a one-parameter family of states associated with orbits generated 
by a Hermitian Hamiltonian ${\hat H}$, there are many 
other situations of interests in quantum theory where the notion of a statistical manifold 
${\mathfrak M}$ plays an important role. For example, the 
parameters $\{\theta^j\}$ may represent the coordinates for atomic coherent states, in which case 
(\ref{eq:16}) determines the metric of the coherent-state manifold (see \cite{BG1} and references 
cited therein for a detailed calculation of the geometry of coherent states). Alternatively, and this is 
the case of interest here, if $|\xi\rangle$ represents an eigenstate of a Hamiltonian ${\hat H}$ such 
that some of the parameters in ${\hat H}$ can be adjusted, then we obtain another example of a 
statistical manifold ${\mathfrak M}$. The geometry of ${\mathfrak M}$ can then exhibit nontrivial 
behaviour for systems describing quantum phase transitions. 

In the context of an information-geometric analysis of quantum phase transitions it has been 
pointed out in \cite{ZGC} (see also \cite{Zhu,Hamma} for a closely related analysis) that 
perturbation analysis can 
be effective in gaining insights into the properties of the metric (\ref{eq:16}). Traditionally, in the 
literature on quantum phase transitions there is a lot of focus on the behaviour of the ground state; 
however, transitions can occur in a multitude of ways. Here we consider an $n^{\rm th}$ 
eigenstate of a Hermitian Hamiltonian ${\hat H}(\theta)$: 
\begin{eqnarray}
{\hat H}|\phi_n\rangle = E_n |\phi_n\rangle  ,
\label{eq:18}
\end{eqnarray}
where for simplicity of notation we have omitted the $\theta$-dependence of ${\hat H}$, $E_n$, and 
$|\phi_n\rangle$. Assuming that the eigenvalues of ${\hat H}(\theta)$ are nondegenerate we can 
use first-order perturbation theory to deduce that 
\begin{eqnarray}
|\partial_a\phi_n\rangle = \sum_{m\neq n} \frac{\langle \phi_m|\partial_a {\hat H}|\phi_n\rangle}
{E_n-E_m} |\phi_m\rangle . 
\label{eq:19}
\end{eqnarray}
Substituting this in (\ref{eq:16}) we find that 
\begin{eqnarray}
G_{ab} = 4 \sum_{m\neq n} 
\frac{\langle \phi_n|\partial_{(a}{\hat H}|\phi_m\rangle\langle \phi_m|\partial_{b)}
{\hat H}|\phi_n\rangle}{(E_n-E_m)^2} . 
\label{eq:20}
\end{eqnarray}
Observe that the skew-symmetric form obtained from the imaginary part of the expression 
(\ref{eq:20}), without the symmetrisation over indices, is just the Berry curvature form 
appearing in the analysis of geometric phases \cite{Berry}. 

To gain intuition about the metric (\ref{eq:20}), consider the problem of estimating the values of 
the parameters appearing in the Hamiltonian, when the system is prepared in the $n^{\rm th}$ 
eigenstate. In a region where the state $|\phi_n\rangle$ is 
sensitive to the changes of the parameter values, the components of the Fisher-Rao metric 
(\ref{eq:20}) are large, and the estimation can be made accurately. If the system exhibits quantum 
phase transitions where one or more of the eigenvalues approach the level $E_n$, then the metric 
becomes singular. Of course, the perturbation (\ref{eq:19}) is applicable only away from 
degeneracies, and hence in the vicinity of degeneracies higher-order perturbative analysis is 
required to identify detailed properties of the metric geometry. In addition, the metric tensor is not 
invariant under coordinate transformations, hence for a more comprehensive analysis one is 
required to work out an expression for the Ricci scalar. Such an analysis would shed further light on 
the theoretical study of quantum phase transitions. 

An alternative way of interpreting the metric (\ref{eq:20}) has been suggested in \cite{ZGC}, which 
we shall develop further here since it is relevant to information-geometric considerations. For 
$\rd\theta\ll1$, and away from degeneracies, we define the unitary operator according to the 
prescription 
\begin{eqnarray}
{\hat U} = \sum_n |\phi_n(\theta+\rd\theta)\rangle\langle\phi_n(\theta)| . 
\label{eq:21}
\end{eqnarray}
Evidently, ${\hat U}$ transports the state $|\phi_n(\theta)\rangle$ into $|\phi_n(\theta+\rd\theta)
\rangle$. The generators of this evolution are then given by the observables 
\begin{eqnarray}
{\hat X}_a = \ri (\partial_a {\hat U}){\hat U}^{-1} . 
\label{eq:22}
\end{eqnarray}
It is then a short exercise to show that the Fisher-Rao metric is just the covariance matrix for the 
observables ${\hat X}_a$ \cite{ZGC}. 

Now if we let ${\hat\Theta}^a$ denote the unbiased estimator 
for the parameter $\theta^a$, then the two operators ${\hat\Theta}^a$ and ${\hat X}_a$ are 
conjugate to each other. In particular, from the Cram\'er-Rao inequality we find that the covariance 
matrix of ${\hat\Theta}^a$ is bounded below by the reciprocal of the Fisher-Rao metric. Hence the 
operator pair $({\hat\Theta}^a,{\hat X}_a)$ for each $a$ satisfies a Heisenberg-like uncertainty 
relation. As an example, suppose that there is a single control parameter $\theta$ in the 
Hamiltonian, and that 
${\hat\Theta}$ is the unbiased estimator for $\theta$, satisfying $\langle\phi_n(\theta)|{\hat\Theta}
|\phi_n(\theta)\rangle=\theta$. (In general, ${\hat\Theta}$ will not be a self-adjoint operator.) 
Suppose, further, that $\lambda^{-1}{\hat X}$ is the self-adjoint operator generating the shift in 
the parameter $\theta$ so that $\re^{-{\rm i}{\hat X}\epsilon/\lambda}\phi(\theta)=\phi(\theta+
\epsilon)$ for $\epsilon\ll1$. Here, $\lambda$ is a constant such that ${\hat X}\epsilon/\lambda$ 
is dimensionless. In this situation, parameter estimate for $\theta$ is limited by the variance 
lower bound of the form: 
\begin{eqnarray}
\Delta {\Theta}^2 \, \Delta {X}^2 \geq \frac{\lambda^2}{4}, 
\label{eq:x25} 
\end{eqnarray}
where by $\Delta{\Theta}^2$ we mean the variance of ${\hat\Theta}$, and similarly for 
$\Delta{X}^2$. It also follows (setting $\lambda=1$) that 
\begin{eqnarray}
\Delta {X}^2 = \sum_{m\neq n} \frac{\langle \phi_n|{\hat H}'|\phi_m\rangle
\langle \phi_m|{\hat H}'|\phi_n\rangle}{(E_n-E_m)^2} , 
\end{eqnarray}
where we have written ${\hat H}'=\partial_\theta{\hat H}$. We remark that (\ref{eq:x25}) represents 
a new type of uncertainty relation in quantum mechanics that is in principle verifiable in laboratory 
experiments.  

The perturbation analysis indicated above can also be applied to obtain an expression for the 
curvature of a 
curve associated with a one-parameter family of eigenstates $|\phi_n(\theta)\rangle$ of a 
parametric Hamiltonian ${\hat H}(\theta)$. In the one-parameter case (\ref{eq:19}) reduces to 
\begin{eqnarray}
|{\dot\phi}_n\rangle = \sum_{m\neq n} \frac{\langle \phi_m|{\hat H}'|\phi_n\rangle}
{E_n-E_m} |\phi_m\rangle . 
\label{eq:23}
\end{eqnarray} 
Assuming that ${\hat H}(\theta)$ is 
nondegenerate, the second-order term in perturbation series gives 
\begin{eqnarray}
|{\ddot\phi}_n\rangle = 2\sum_{m\neq n} \left[ \sum_{l\neq n} 
\frac{\langle \phi_m|{\hat H}'|\phi_l\rangle\langle \phi_l|{\hat H}'|\phi_n\rangle}
{(E_n-E_m)(E_n-E_l)} - 
\frac{\langle \phi_n|{\hat H}'|\phi_n\rangle\langle \phi_m|{\hat H}'|\phi_n\rangle}
{(E_n-E_m)^2} \right] |\phi_m\rangle , 
\label{eq:24}
\end{eqnarray}
which shows that $\langle\phi_n|{\ddot\phi}_n\rangle =0$. 
In this case the expression for the intrinsic curvature becomes: 
\begin{eqnarray}
{\mathcal K}_n^2 = \frac{\langle{\ddot\phi}_n|{\ddot\phi}_n\rangle}
{\langle{\dot\phi}_n|{\dot\phi}_n\rangle^2} - 
\frac{\langle{\ddot\phi}_n|{\dot\phi}_n\rangle\langle{\dot\phi}_n|{\ddot\phi}_n\rangle}
{\langle{\dot\phi}_n|{\dot\phi}_n\rangle^3} . 
\label{eq:25}
\end{eqnarray}
Substitution of (\ref{eq:23}) and (\ref{eq:24}) in (\ref{eq:25}) then gives the expression for the 
curvature, which, in turn, can be used (cf. \cite{BH2}) to derive a higher-order correction to \
the uncertainty lower bound (\ref{eq:x25}).

%%%%%%%%%%%%%%%%%%%%%%%%%%%%%%%%%%%%%%%%%%%%%%%%%%%%%%%%%%%%

\section{Information geometry for complex Hamiltonians}
\label{sec:5}

We now wish to examine the statistical manifold ${\mathfrak M}$ associated with eigenstates of 
a complex Hamiltonian ${\hat K}$ for which ${\hat K}^\dagger\neq{\hat K}$. Complex Hamiltonians 
are traditionally used to describe decay and scattering phenomena 
\cite{Maha69,SW,Datt90b,Okol03,Moiseyev2}. They are also used in the 
context of open systems as effective Hamiltonians. More recently, complex Hamiltonians that 
fulfil certain antilinear symmetry have attracted a lot of attention, owing to the facts that such 
Hamiltonians may possess entirely real eigenvalues, and that depending on the parameter values 
in the Hamiltonian there can be a phase transition where a pair of real eigenvalues degenerates 
and turns into a complex conjugate pair 
\cite{DC1,NM,AM,DC2,DC3,SR,TK,SR2,TK2,BD,BG2,CMB}. 
As indicated above, such a critical point is where the associated 
eigenstates also coalesce, thus constituting an example of an exceptional point. Here we are 
interested in the geometry of the statistical manifold ${\mathfrak M}$ associated with such a 
Hamiltonian exhibiting one or more phase transitions. 

To proceed, let ${\hat K}={\hat H}-\ri {\hat{\mathit\Gamma}}$, where ${\hat H}^\dagger={\hat H}$ 
and ${\hat{\mathit\Gamma}}^\dagger={\hat{\mathit\Gamma}}$, be a complex Hamiltonian with 
eigenstates $\{|\phi_n\rangle\}$ and nondegenerate eigenvalues $\{\kappa_n\}$: 
\begin{eqnarray}
{\hat K}|\phi_n\rangle = \kappa_n|\phi_n\rangle \quad {\rm and} \quad \langle\phi_n| 
{\hat K}^\dagger = {\bar\kappa}_n\langle\phi_n| . 
\label{eq:26}
\end{eqnarray} 
Additionally, it will be convenient to introduce eigenstates of the adjoint matrix 
${\hat K}^\dagger$: 
\begin{eqnarray}
{\hat K}^\dagger|\chi_n\rangle = {\bar\kappa}_n|\chi_n\rangle \quad {\rm and} \quad 
\langle\chi_n| {\hat K} = \kappa_n\langle\chi_n| . 
\label{eq:27}
\end{eqnarray} 
The reason for introducing the additional states $\{|\chi_n\rangle\}$ is because the eigenstates 
$\{|\phi_n\rangle\}$ of ${\hat K}$ are in general not orthogonal, and hence conventional projection 
techniques 
so commonly used in many calculations of quantum mechanics, in particular, in perturbation theory, 
are not effective when dealing with the eigenstates of a complex Hamiltonian 
\cite{Pell,Cloizeaux,more,Curtright,Moiseyev2}. 
With the introduction of the states $\{|\chi_n\rangle\}$, however, we have the relations: 
\begin{eqnarray}
\langle\chi_n|\phi_m\rangle = \delta_{nm}\langle\chi_n|\phi_n\rangle \quad {\rm and} \quad 
\sum_n \frac{|\phi_n\rangle \langle\chi_n|}{\langle\chi_n|\phi_n\rangle} = {\mathds 1} ,
\label{eq:30}
\end{eqnarray}
which hold in finite dimensions away from degeneracies. 

With the use of the biorthogonal states the notion of an associated state can be introduced: For 
an arbitrary state $|\psi\rangle$, we define the associated state $|{\tilde\psi}\rangle$ according to 
the following relations: 
\begin{eqnarray}
|\psi\rangle = \sum_n c_n |\phi_n\rangle \quad \Leftrightarrow \quad 
\langle{\tilde\psi}| = \sum_n {\bar c}_n \langle\chi_n| . 
\label{eq:48}
\end{eqnarray}
We shall let (\ref{eq:48}) determine the duality relation on the state space. Additionally, for 
convenience we assume that $\langle\chi_n|\phi_n\rangle=1$ holds for all $n$. Under this 
convention the states are no longer normalised, i.e. $\langle\psi|\psi\rangle>1$, 
but we can assume that 
\begin{eqnarray}
\langle{\tilde\psi}|\psi\rangle=\sum_n{\bar c}_nc_n=1. 
\label{eq:49}
\end{eqnarray}

At an exceptional point, however, the convention $\langle\chi_{EP}|\phi_{EP}\rangle=1$ breaks down 
for the following reason. Suppose that the two eigenstates $|\phi_k\rangle$ and $|\phi_l\rangle$ 
`meet' at $|\phi_{EP}\rangle$. Evidently, the biorthogonality condition implies that 
$\langle\chi_{l}|\phi_{k}\rangle=0$ and $\langle\chi_{k}|\phi_{k}\rangle\neq0$, but $\langle\chi_{l}|$ 
and $\langle\chi_{k}|$ will both approach $\langle\chi_{EP}|$ so that we have $\langle\chi_{EP}
|\phi_{EP}\rangle=0$. This feature is often referred to as `self-orthogonality' in the literature. To 
complete the basis for the eigenspace belonging to the degenerate eigenstate one needs to 
introduce associated eigenvactors, or so-called \textit{Jordan vectors}. We will return to this issue 
in the discussion of exceptional points in the section to follow, but for now we assume that the states are away from degeneracies.

Away from exceptional points, and based on the convention that $\langle\chi_n|\phi_n\rangle=1$, 
the overlap distance $s$ between 
the two states $|\xi\rangle$ and $|\eta\rangle$ is now given by the expression: 
\begin{eqnarray}
\cos^2 {\textstyle\frac{1}{2}} s = \frac{\langle{\tilde\xi}|\eta\rangle\langle{\tilde\eta}|\xi\rangle}
{\langle{\tilde\xi}|\xi\rangle\langle{\tilde\eta}|\eta\rangle} . 
\label{eq:51}
\end{eqnarray}
In particular, if 
$|\eta\rangle=|\xi\rangle+|\rd\xi\rangle$ is a neighbouring state to $|\xi\rangle$, then expanding 
(\ref{eq:51}) and retaining terms of quadratic order, we obtain the following form of the 
Fubini-Study line element
\begin{eqnarray}
\rd s^2 = 4 \frac{\langle{\tilde\xi}|\xi\rangle\langle{\widetilde{\rd\xi}}|\rd\xi\rangle - 
\langle{\tilde\xi}|\rd\xi\rangle\langle{\widetilde{\rd\xi}}|\xi\rangle}
{\langle{\tilde\xi}|\xi\rangle^2} . 
\label{eq:52}
\end{eqnarray}
We remark that an analogous expression for the metric appears in \cite{CZ}, however, 
(\ref{eq:52}) is different from the metric obtained in \cite{CZ} since we have chosen a 
different definition for an associated state $\langle{\tilde\xi}|$. 

With the alternative expression (\ref{eq:52}) for the Fubini-Study metric at hand we are in the 
position to investigate the metric geometry of eigenstates of complex Hamiltonians. To begin, 
recall that for the identification of the local metric geometry of the statistical manifold 
${\mathfrak M}$ associated with an eigenstate $|\phi_n\rangle$ of a Hamiltonian ${\hat K}$ we 
need to determine the perturbation $|\partial_a\phi_n\rangle\rd\theta^a$ of the state associated 
with a small change in the parameter values. For this purpose we shall follow closely the approach 
of \cite{SW}. Specifically, with the convention $\langle\chi_n|\phi_n\rangle=1$ and the help of the 
biorthogonal states $(\{|\phi_n \rangle\},\{|\chi_n\rangle\})$, we consider the perturbation of an 
eigenstate away from degeneracies. Then the eigenvalues and eigenvectors can be expanded in 
a Taylor series in the perturbation parameter, much as in the Hermitian case. Taylor expand the 
Hamiltonian ${\hat K}(\theta)$, the eigenstate $|\phi_n(\theta)\rangle$, and the eigenvalue 
$\kappa_n(\theta)$ at $\theta$ in the eigenvalue equation, we obtain 
\begin{eqnarray}
({\hat K}+\partial_a{\hat K} \rd \theta^a +\cdots) (|\phi_n\rangle+|\partial_a\phi_n\rangle\rd\theta^a 
+\cdots) = (\kappa_n+\partial_a\kappa_n\rd\theta^a+\cdots) 
(|\phi_n\rangle+|\partial_a\phi_n\rangle\rd\theta^a \cdots)  ,
\end{eqnarray}
where we have omitted explicit $\theta$ dependencies. Equating the terms linear in $\rd\theta$ we 
find 
\begin{eqnarray}
({\hat K} - \kappa_n) |\partial_a\phi_n\rangle = 
\partial_a\kappa_n |\phi_n\rangle - \partial_a{\hat K} |\phi_n\rangle  . 
\label{eq:39}
\end{eqnarray}

So far the result is identical to that for a Hermitian Hamiltonian. However, the lack of orthogonality 
of the eigenstates prevents us from using the projector ${\hat\Phi}_m=|\phi_m\rangle\langle\phi_m|$ 
to further simplify the expression. Nevertheless, if we multiply ${\hat\Pi}_m=|\phi_m\rangle\langle
\chi_m|$ from the left and rearrange terms we find 
\begin{eqnarray}
(\kappa_m - \kappa_n) {\hat\Pi}_m  |\partial_a\phi_n\rangle = (\partial_a\kappa_n) \delta_{mn} 
|\phi_m\rangle-\langle \chi_m|\partial_a {\hat K}|\phi_n\rangle |\phi_m\rangle . 
\label{eq:40}
\end{eqnarray}
For $n=m$ we are led to the expression (cf. \cite{more}): 
\begin{eqnarray}
\partial_a\kappa_n = \langle \chi_n|\partial_a {\hat K}|\phi_n\rangle . 
\label{eq:41}
\end{eqnarray}
To obtain an expression for $|\partial_k\phi_n\rangle$, in \cite{SW} the operator 
$({\hat K} - \kappa_n{\mathds 1})^{-1}$ is applied from the left in (\ref{eq:39}). This approach, 
however, is problematic on account of the fact that $({\hat K} - \kappa_n{\mathds 1})$ is degenerate 
and thus not invertible. The result of \cite{SW} can nevertheless be justified if we make the assumption 
that the perturbation vector $|\partial_a\phi_n\rangle \rd\theta^a$ is orthogonal to the dual vector 
$|\chi_n\rangle$. With this assumption, which turns out to be the correct one, for $n\neq m$ we 
divide both sides of (\ref{eq:40}) by $\kappa_m-\kappa_n$ and sum over $m\neq n$ to obtain
\begin{eqnarray}
|\partial_a\phi_n\rangle = \sum_{m\neq n} \frac{\langle \chi_m|\partial_a {\hat K}|\phi_n\rangle}
{\kappa_n-\kappa_m} |\phi_m\rangle ,
\label{eq:42}
\end{eqnarray}
where we have made use of the condition $\langle\chi_m|\rd\phi_m\rangle=0$. 

The perturbation term (\ref{eq:42}) formally resembles the expression (\ref{eq:19}) of its Hermitian 
counterpart. However, there are important differences, including the fact that the perturbation is 
not orthogonal to the state $|\phi_n\rangle$, i.e. $\langle\phi_n|\partial_a\phi_n\rangle\neq0$, but rather $\langle\chi_n|\partial_a\phi_n\rangle=0$. It follows 
that under this assumption the perturbation will necessarily change the overall complex phase of the 
eigenstate. This is nevertheless natural under the geometry of the state space formulated from 
(\ref{eq:52}). 

The metric geometry of the parameter space can now be determined if we substitute (\ref{eq:42}) 
in (\ref{eq:52}): 
\begin{eqnarray}
G_{ab} = 4\sum_{m\neq n} 
\frac{\langle\chi_m|\partial_{(a}{\hat K}|\phi_n\rangle\langle\phi_n|\partial_{b)}{\hat K}|\chi_{m}\rangle}
{({\bar\kappa}_n-{\bar\kappa}_{m})(\kappa_n-\kappa_m)} . 
\label{eq:60}
\end{eqnarray}
With the expression (\ref{eq:60}) at hand we are able to investigate the geometry of the statistical 
manifold associated with eigenstates of complex Hamiltonians, away from degeneracies. 
Incidentally, this expression for the metric is in line with the analysis of geometric phases 
associated with the eigenstates of complex Hamiltonians \cite{GW,MKS}. Since the perturbative 
result (\ref{eq:60}) is only valid away from degeneracies, in the next section we shall investigate 
the generic behaviour close to the exceptional point by employing a more refined perturbative 
technique.

%%%%%%%%%%%%%%%%%%%%%%%%%%%%%%%%%%%%%%%%%%%%%%%%%%%%%%%%%%%%

\section{Geometry close to exceptional points}
\label{sec:6}

In the case of a Hermitian Hamiltonian, the first-order perturbation used to derive expression 
(\ref{eq:20}) for the metric breaks down near degeneracies, and one has to consider higher-order 
perturbations. In the case of a complex Hamiltonian, the situation is more severe on account of 
the fact that the Rayleigh-Schr\"odinger perturbation theory breaks down altogether in the vicinities 
of exceptional points. Nevertheless, for a given Hamiltonian one can expand the 
eigenstates and eigenvalues in the form of Newton-Puiseux series in order to identify the metric 
geometry close to exceptional points (see, for example, \cite{GRS,GGKN} for effective use of the 
Newton-Puiseux expansion for the investigation of properties of the eigenstates of complex 
Hamiltonians in the vicinities of exceptional points; see also \cite{arnold,SM} for a more general 
discussion on related mathematical ideas). This line of investigation therefore leads to a new 
application of information geometry in the sensitivity analysis of physical systems characterised 
by Hermitian or more generally complex Hamiltonians (we remark that properties of exceptional 
points of higher order where more than two eigenstates coalesce can be quite intricate; see, e.g., 
\cite{DG,HC}). 

Let us illustrate how such an analysis can be applied to deduce the nature of geometric 
singularities close to exceptional points. For more details on perturbation theory around 
exceptional points see, e.g., \cite{SM} and references cited therein. As indicated above, at 
an exceptional point two or more eigenvalues and the corresponding eigenstates coalesce, 
that is, the Hamiltonian is not diagonalisable. Here we consider the most common case, 
where two eigenvalues and the corresponding eigenstates coalesce. At such an exceptional 
point there is a two-fold degenerate eigenvalue $\kappa_{EP}$ and a single eigenvector 
$|\phi_{EP}\rangle$, which is orthogonal to the corresponding left eigenvector: 
$\langle\chi_{EP}|\phi_{EP}\rangle=0$. However, one can define an associated vector, the 
so-called Jordan vector, denoted $|\phi_{EP}^{J}\rangle$, fulfilling the relation
\begin{eqnarray}
\label{def_Jordan}
{\hat K}|\phi_{EP}^J\rangle=\kappa_{EP}|\phi_{EP}^J\rangle+|\phi_{EP}\rangle.
\end{eqnarray}
Similarly the left Jordan vector can be defined according to the relation 
\begin{eqnarray}
\label{def_Jordan2}
{\hat K}^\dagger|\chi_{EP}^J\rangle={\bar\kappa}_{EP}|\chi_{EP}^J\rangle+|\chi_{EP}\rangle .
\end{eqnarray} 
The Jordan vector $|\phi_{EP}^J\rangle$ and the eigenvector $|\phi_{EP}\rangle$ 
span the two-dimensional eigenspace corresponding to the degenerate eigenvalue 
$\kappa_{EP}$. Note that the Jordan vector is not uniquely defined by equation (\ref{def_Jordan}). 
However, the ambiguity can be removed by choosing appropriate normalisation conditions \cite{SM}. In fact, it 
will be convenient to normalise the states such that 
\begin{eqnarray}
\langle \chi_{EP}|\phi_{EP}^J\rangle=\langle \chi_{EP}^J|\phi_{EP}\rangle=1,
\end{eqnarray}
and that 
\begin{eqnarray}
\label{norm_EP}
\langle \chi_{EP}^J|\phi_{EP}^J\rangle=0.
\end{eqnarray}

As already indicated, conventional Rayleigh-Schr\"odinger perturbation theory breaks down 
around an exceptional point, and in general the eigenvalues and eigenvectors are not analytic 
functions of the perturbation parameter. That is, they cannot be expanded in a Taylor series. 
In the general case they can nevertheless be expanded into a power series with broken rational 
exponents, which is known as a Puiseux series. While in general one has to distinguish different 
cases of perturbation behavious \cite{Ma}, the most common, generic behaviour around an 
exceptional point at which two eigenvectors coalesce is that the eigenvalues and eigenvectors 
can be expanded in a power series with half-integral exponents.

Let $\epsilon\ll1$ denote a small 
perturbation parameter that measures the deviation away from the exceptional point. 
Expanding the Hamiltonian, the eigenvalues and eigenvectors in lowest order in $\epsilon$ in the eigenvalue equation yields
\begin{equation}
({\hat K_{EP}}+\epsilon{\hat K}' +\cdots) (|\phi_{EP}\rangle+|\phi'\rangle\epsilon^{\frac{1}{2}} 
+\cdots) = (\kappa_{EP}+\kappa'\epsilon^{\frac{1}{2}}+\cdots) 
(|\phi_{EP}\rangle+|\phi'\rangle\epsilon^{\frac{1}{2}} +\cdots)
\end{equation}
Equating terms corresponding to different powers of $\epsilon$ and using equations (\ref{def_Jordan})-(\ref{norm_EP}) we find that the two eigenstates $|\phi_\pm\rangle$ can 
be expanded in the vicinity of an exceptional point in the form: 
\begin{eqnarray}
\label{EP_pert_phi}
|\phi_\pm\rangle=n\left(|\phi_{EP}\rangle+\kappa'_\pm\,\epsilon^{\frac{1}{2}}
\,|\phi_{EP}^J\rangle+O(\epsilon)\right), 
\end{eqnarray}
where 
\begin{eqnarray}
\kappa'_\pm=\pm\sqrt{\langle\chi_{EP}|{\hat K}'|\phi_{EP}\rangle}.
\end{eqnarray}
A perturbative expression similar to (\ref{EP_pert_phi}) holds for the left eigenvector. The 
resulting left and right eigenvectors are automatically orthogonal, however, they are only 
defined up to a normalisation constant $n$. It is convenient to normalise these vectors 
according to the usual biorthogonal convention away from the exceptional point: 
$\langle\chi_\pm|\phi_\pm\rangle=1$. From this we find
\begin{eqnarray}
|\phi_\pm\rangle\approx\frac{1}{\sqrt{2\kappa'}\epsilon^{1/4}}\left(|\phi_{EP}\rangle+
\kappa'\,\epsilon^{\frac{1}{2}}\,|\phi_{EP}^J\rangle\right), \quad 
\langle\chi_\pm|\approx\frac{1}{\sqrt{2\kappa'}\epsilon^{1/4}}\left(\langle\chi_{EP}|
+\kappa'\,\epsilon^{\frac{1}{2}}\,\langle\chi_{EP}^J|\right). 
\end{eqnarray}
A calculation then shows that
\begin{eqnarray}
|\rd\phi_+\rangle = \frac{1}{4\sqrt{\kappa'}}\left(-\epsilon^{-\frac{5}{4}}|\phi_{EP}\rangle
+\kappa'\,\epsilon^{-\frac{3}{4}}\,|\phi_{EP}^J\rangle\right)\rd\epsilon 
= \frac{1}{4\epsilon}\,|\phi_-\rangle \rd \epsilon ,
\label{eq:w47}
\end{eqnarray}
and hence that 
\begin{eqnarray}
\langle \widetilde{\rd\phi_+}|=\frac{1}{4\epsilon}\langle\chi_-| \rd \epsilon. 
\label{eq:w48}
\end{eqnarray}
From (\ref{eq:w47}) and (\ref{eq:w48}) we thus find the expression of the metric 
close to an exceptional point of second order where two eigenstates coalesce:
\begin{eqnarray}
\label{G_EP_pert}
G=\frac{1}{4\epsilon^2}, 
\end{eqnarray}
on account of (\ref{eq:52}). It should be remarked that the result (\ref{G_EP_pert}) is generic, 
i.e. it is independent of the model. It can therefore be viewed as providing the scaling property 
of the metric close to an exceptional point of second order, in a manner analogous to the 
scaling behaviour of the metric near critical points in statistical mechanics of phase 
transitions \cite{BN}.

\section{Discussion}
\label{sec:7}

We conclude by remarking that although in the foregoing material we have placed some emphasis 
on perturbative analysis for the geometry surrounding exceptional points so as to obtain generic 
expressions for the metric, if a model is specified, then typically there is no need for evoking the 
perturbative approach since the metric can be computed exactly. As an example, take the 
$2\times2$ Hamiltonian ${\hat K}={\hat\sigma}_x-\ri\gamma{\hat\sigma}_z$. This Hamiltonian is 
PT symmetric, and has real eigenvalues in the region $\gamma^2<1$ where the eigenstates are 
also PT symmetric. Specifically, the eigenstates of ${\hat K}$ and ${\hat K}^\dagger$  are given by 
\begin{eqnarray}
|\phi_\pm\rangle = n_\pm \left( \begin{array}{c} 1 \\ \ri\gamma\pm\sqrt{1-\gamma^2} 
\end{array} \right) , \qquad 
|\chi_\pm\rangle = n_\mp \left( \begin{array}{c} 1 \\ -\ri\gamma\pm\sqrt{1-\gamma^2} 
\end{array} \right), 
\end{eqnarray}
where $n_\pm^2=(1\mp\ri\gamma/\sqrt{1-\gamma^2})/2$. A straightforward calculation then 
shows that the information metric associated with the curve, say, $|\phi_+(\gamma)\rangle$, is 
given by 
\begin{eqnarray}
G=\frac{1}{(1-\gamma^2)^{2}}, 
\label{eq:64}
\end{eqnarray} 
on account of the relations: 
\begin{eqnarray}
|\rd\phi_+\rangle=-\frac{\ri\, \rd\gamma}{2(1-\gamma^2)}|\phi_-\rangle, \qquad 
\langle\widetilde{\rd\phi_+}|=\frac{\ri\,\rd\gamma}{2(1-\gamma^2)}\langle\chi_-| . 
\end{eqnarray} 
The nonperturbative expression in (\ref{eq:64}) shows exactly how the metric diverges as one 
approaches the critical point $\gamma_c=1$. It can be easily verified that (\ref{eq:64}) also holds 
when the singularity is approached from the region $\gamma^2>1$. To compare this exact result 
with the perturbative analysis presented in the previous section, let us write 
$\gamma=\gamma_c-\epsilon$. Then we find 
\begin{eqnarray}
G=\frac{1}{4\epsilon^2+4\epsilon^3+\epsilon^4} = \frac{1}{4\epsilon^2}\left( 1 - \epsilon 
+ \frac{3}{4}\epsilon^2 - \cdots \right) ,  
\end{eqnarray}
thus recovering the perturbative result (\ref{G_EP_pert}) in leading order of $\epsilon$.

More generally, any curve of the form 
$|\psi(\gamma)\rangle=c_+|\phi_+(\gamma)\rangle+c_-|\phi_-(\gamma)\rangle$ with fixed 
coefficients $c_\pm$ in this system possesses the metric (\ref{eq:64}) and will exhibit a curvature 
singularity at $\gamma=1$. In the region $\gamma^2\gg1$, on the other hand, we have $G\ll1$, 
and thus estimation of the parameter $\gamma$ becomes unfeasible. 

%%%%%%%%%%%%%%%%%%%%%%%%%%%%%%%%%%%%%%%%%%%%%%%%%%%%%%%%%%%%

\section*{Acknowledgements}

The authors thank participants of 12$^{\rm th}$ International Conference on Non-Hermitian 
Operators in Quantum Physics, Istanbul, July 2013, for stimulating discussion. 

%==========================================================
%==========================================================
% Back Matter (References and Notes)
%----------------------------------------------------------
% Style and layout of the references
\bibliographystyle{mdpi}
\makeatletter
\renewcommand\@biblabel[1]{#1. }
\makeatother
%----------------------------------------------------------
% Use the following option to include external BibTeX files:
%\bibliography{template}

\begin{thebibliography}{1}


\bibitem{DCB2} 
Brody,~D.~C., Hook,~D.~W. \& Hughston,~L.~P.
Quantum phase transitions without thermodynamic limits.
{\em Proceedings of the Royal Society London} A\textbf{463}, 2021-2030 (2007).

\bibitem{YL} 
Yang,~C.~N. \& Lee,~T.~D. 
Statistical theory of equations of state and phase transitions. I. Theory of condensation. 
{\em Physical Review} \textbf{87}, 404-409 (1952). 

\bibitem{LY} 
Lee,~T.~D. \& Yang,~C.~N.  
Statistical theory of equations of state and phase transitions. II. Lattice gas and Ising model. 
{\em Physical Review} \textbf{87}, 410-419 (1952). 

\bibitem{BE} 
Blythe,~R.~A. \& Evans,~M.~R. 
The Lee-Yang theory of equilibrium and nonequilibrium phase transitions. 
{\em Brazilian Journal of Physics} \textbf{33}, 464-475 (2003). 

\bibitem{CHM} 
Cejnar,~P., Heinze,~S. \& Macek,~M. 
Coulomb analogy for non-Hermitian degeneracies near quantum phase transitions. 
{\em Physical Review Letters} \textbf{99}, 100601 (2007).

\bibitem{Heiss} 
Heiss,~W.~D. 
The physics of exceptional points. 
{\em Journal of Physics} A\textbf{45}, 444016 (2012). 

\bibitem{BH} 
Brody,~D.~C. \& Hook,~D.~W.  
Information geometry in vapour-liquid equilibrium. 
{\em Journal of Physics} A\textbf{42}, 023001 (2009). 

\bibitem{ZGC} 
Zanardi,~P., Giorda,~P. \& Cozzini,~M.   
Information-theoretic differential geometry of quantum phase transitions.  
{\em Physical Review Letters} \textbf{99}, 100603 (2007). 

\bibitem{Panc55}
Pancharatnam,~S.  
The propagation of light in absorbing biaxial crystals. II. Experimental. 
{\em Proceedings of the Indian Academy of Science} A\textbf{42} 235-248 (1955)

\bibitem{Richter} 
Dembowski,~C., Gr\"af,~H.~D., Harney,~H.~L., Heine,~H.~L., Heiss,~W.~D., 
Rehfeld,~H. \& Richter,~A. 
Experimental observation of the topological structure of exceptional points. 
{\em Physical Review Letters} \textbf{86}, 787-790 (2001). 

\bibitem{Richter2} 
Dembowski,~C., Dietz,~B., Gr\"af,~H.~D., Harney,~H.~L., Heine,~H.~L., Heiss,~W.~D. 
\& Richter,~A. 
Observation of a chiral state in a microwave cavity. 
{\em Physical Review Letters} \textbf{90}, 034101 (2003). 

\bibitem{Lee09}
Lee,~S.~B., Yang,~J., Moon,~S., Lee,~S.~Y., Shim,~J.~B., Kim,~S.~W., Lee,~J.~H. \& An,~K. Observation of an exceptional point in a chaotic optical microcavity. 
{\em Physical Review Letters} \textbf{103}, 134101 (2009)

\bibitem{BB} 
Bender,~C.~M. \& Boettcher,~S. 
Real spectra in non-Hermitian Hamiltonians having PT symmetry. 
{\em Physical Review Letters} \textbf{80}, 5243-5246 (1998).

\bibitem{DC1} 
Makris,~K.~G., El-Ganainy,~R., Christodoulides,~D.~N. \& Musslimani~Z.~H.
Beam dynamics in PT symmetric optical lattices.  
{\em Physical Review Letters} \textbf{100}, 103904 (2008).

\bibitem{NM} 
Klaiman,~S., G\"unther,~U. \& Moiseyev,~N.  
Visualization of branch points in PT-symmetric waveguides.  
{\em Physical Review Letters} \textbf{101}, 080402 (2008).

\bibitem{AM} 
Mostafazadeh,~A. 
Spectral singularities of complex scattering potentials and infinite reflection 
and transmission coefficients at real energies. 
{\em Physical Review Letters} \textbf{102}, 220402 (2009).

\bibitem{DC2} 
Guo,~A., Salamo,~G.~J., Duchesne,~D., Morandotti,~R., Volatier-Ravat,~M., 
Aimez,~V., Siviloglou,~G.~A. \& Christodoulides,~D.~N.  
Observation of PT-symmetry breaking in complex optical potentials. 
{\em Physical Review Letters} \textbf{103}, 093902 (2009).

\bibitem{DC3} 
R\"uter,~C.~E., Makris,~K.~G., El-Ganainy,~R., Christodoulides,~D.~N. \& Kip,~D.  
Observation of parityÐtime symmetry in optics. 
{\em Nature Physics} \textbf{6}, 192-195 (2010). 

\bibitem{SR} 
Ge,~L., Chong,~Y.~D., Rotter,~S., T\"ureci,~H.~E. \& Stone,~A.~D.  
Unconventional modes in lasers with spatially varying gain and loss. 
{\em Physical Review} A\textbf{84}, 023820 (2011). 

\bibitem{TK} 
Schindler,~J., Li,~A., Zheng,~M.~C., Ellis,~F.~M. \& Kottos,~T. 
Experimental study of active LRC circuits with PT symmetries. 
{\em Physical Review} A\textbf{84}, 040101 (2011).

\bibitem{SR2} 
Liertzer,~M., Ge,~L., Cerjan,~A., Stone,~A.~D., T\"ureci,~H.~E. \& Rotter,~S. 
Pump-induced exceptional points in lasers. 
{\em Physical Review Letters} \textbf{108}, 173901 (2012). 

\bibitem{TK2} 
Ramezani,~H., Schindler,~J., Ellis,~F.~M., G\"unther,~U. \& Kottos,~T.  
Bypassing the bandwidth theorem with PT symmetry. 
{\em Physical Review} A\textbf{85}, 062122 (2012). 

\bibitem{BD} 
Bittner,~S., Dietz,~B., G\"unther,~U., Harney,~H.~L., Miski-Oglu,~M.,  
Richter,~A. \& Sch\"afer,~F.  
PT symmetry and spontaneous symmetry breaking in a microwave billiard. 
{\em Physical Review Letters} \textbf{108}, 024101 (2012). 

\bibitem{BG2} 
Brody,~D.~C. \& Graefe,~E.~M. 
Mixed-state evolution in the presence of gain and loss. 
{\em Physical Review Letters} \textbf{109}, 230405 (2012).

\bibitem{CMB} 
Bender,~C.~M., Berntson,~B.~K., Parker,~D. \& Samuel,~E. 
Observation of PT phase transition in a simple mechanical system 
{\em American Journal of Physics} \textbf{81}, 173-179 (2013). 

\bibitem{Rao} 
Rao,~C.~R. 
Information and the accuracy attainable in the estimation of statistical parameters. 
{\em Bulletin of the Calcutta Mathematical Society} \textbf{37}, 81-91 (1945).

\bibitem{fisher} 
Fisher,~R.~A.  
Theory of statistical estimation. 
{\em Proceedings of the Cambridge Philosophical Society} \textbf{22}, 700-725 (1925).

\bibitem{BH0} 
Brody,~D.~C. \& Hughston,~L.~P.  
Geometry of quantum statistical inference. 
{\em Physical Review Letters} \textbf{77}, 2851-2854 (1996). 

\bibitem{Mandelbrot} 
Mandelbrot,~B. 
The role of sufficiency and of estimation in thermodynamics. 
{\em Annals of Mathematical Statistics} \textbf{33}, 1021-1038 (1962). 

\bibitem{BH3} 
Brody,~D.~C. \& Hughston,~L.~P.  
Geometrisation of statistical mechanics.  
{\em Proceedings of the Royal Society London} A\textbf{455}, 1683-1715 (1999).

\bibitem{AA} 
Anandan,~J. \& Aharonov,~Y. 
Geometry of quantum evolution. 
{\em Physical Review Letters} \textbf{65}, 1697-1700 (1990). 

\bibitem{holevo} 
Holevo,~A. 
{\em Probabilistic and Statistical Aspects of Quantum Theory}. 
(Pisa: Edizioni della Normale, 2011).

\bibitem{BG1} 
Brody,~D.~C. \& Graefe,~E.~M. 
Coherent states and rational surfaces. 
{\em Journal of Physics} A\textbf{43}, 255205 (2010).

\bibitem{Zhu} 
Zhu,~S.~L. 
Scaling of geometric phases close to the quantum phase transition in the XY spin chain.  
{\em Physical Review Letters} \textbf{96}, 077206 (2006). 

\bibitem{Hamma} 
Hamma,~A. 
Berry phases and quantum phases transitions.  
arXiv:quant-ph/0602091 (2006). 

\bibitem{Berry}
Berry,~M.~V. 
Quantal phase factors accompanying adiabatic changes. 
{\em Proceedings of the Royal Society London} A\textbf{392}, 45-57 (1984). 

\bibitem{BH2} 
Brody,~D.~C. \& Hughston,~L.~P.  
Statistical geometry in quantum mechanics.  
{\em Proceedings of the Royal Society London} A\textbf{454}, 2445-2475 (1998).

\bibitem{Maha69}
Mahaux,~C. \& Weidenm\"uller,~H.~A. 
{\em Shell Model Approach to Nuclear Reactions}. 
(Amsterdam: North Holland Publishing Company, 1969). 
  
\bibitem{SW} 
Sternheim,~M.~M. \& Walker,~J.~F. 
Non-Hermitian Hamiltonians, decaying states, and perturbation theory. 
{\em Physical Review} C\textbf{6}, 114-121 (1972).

\bibitem{Datt90b}
Dattoli,~G., Torre,~A. \& Mignani,~R.  
Non-Hermitian evolution of two-level quantum systems. 
{\em Physical Review} A\textbf{42} 1467-1475 (1990). 

\bibitem{Okol03}
Oko{\l}owicz,~J., P{\l}oszajczak,~M. \& Rotter,~I.  
Dynamics of quantum systems embedded in a continuum. 
{\em Physics Report} \textbf{374}  271-383 (2003).

\bibitem{Moiseyev2} 
Moiseyev,~N.  
{\em Non-Hermitian Quantum Mechanics}.  
(Cambridge: Cambridge University Press, 2011).

\bibitem{Pell} 
Pell,~A.~J. 
Biorthogonal systems of functions.  
{\em Transactions of the American Mathematical Society} \textbf{12}, 135-164 (1911).

\bibitem{Cloizeaux} 
des~Cloizeaux,~J.   
Extension d'une formule de Lagrange \`a des probl\`emes de valeurs propres. 
{\em Nuclear Physics} \textbf{20}, 321-346 (1960).

\bibitem{more} 
More,~R.M.  
Theory of decaying states. 
{\em Physical Review} A\textbf{4}, 1782-1790 (1971).

\bibitem{Curtright} 
Curtright,~T. \& Mezincescu,~L.  
Biorthogonal quantum systems. 
{\em Journal of Mathematical Physics} \textbf{48}, 092106 (2007).

\bibitem{CZ}
Cui,~X.~D. \& Zheng,~Y.  
Geometric phases in non-Hermitian quantum mechanics.  
{\em Physical Review} A\textbf{86}, 064104 (2012).

\bibitem{GW} 
Garrison,~J.~C. \& Wright,~E.~M. 
Complex geometrical phases for dissipative systems. 
{\em Physics Letters} A\textbf{128}, 177-181 (1988).

\bibitem{MKS} 
Mailybaev,~A.~A., Kirillov,~O.~N. \& Seyranian,~A.~P. 
Geometric phase around exceptional points. 
{\em Physical Review} A\textbf{72}, 014104 (2005).

\bibitem{GRS} 
G\"unther,~U., Rotter,~I. \& Samsonov,~B.~F.  
Projective Hilbert space structures at exceptional points. 
{\em Journal of Physics} A\textbf{40}, 8815-8833 (2007).

\bibitem{GGKN} 
Graefe,~E.~M., G\"unther,~U., Korsch,~H.~J. \& Niederle,~A.~E. 
A non-Hermitian symmetric Bose-Hubbard model: eigenvalue rings from unfolding 
higher-order exceptional points. 
{\em Journal of Physics} A\textbf{41}, 255206 (2008). 

\bibitem{arnold} 
Arnold,~V.~I.  
On matrices depending on parameters. 
{\em Russian Mathematical Survey} \textbf{26}, 29-43 (1971). 

\bibitem{SM} 
Seyranian,~A.~P. \& Mailybaev,~A.~A. 
{\em Multiparameter Stability Theory with Mechanical Applications}. 
(Signapore: World Scientific, 2003). 

\bibitem{DG} 
Demange,~G. \& Graefe,~E.~M.  
Signatures of three coalescing eigenfunctions.  
{\em Journal of Physics} A\textbf{45}, 025303 (2012). 

\bibitem{HC} 
Dast,~D., Haag,~D., Cartarius,~H., Main,~J. \& Wunner,~G. 
Eigenvalue structure of a Bose-Einstein condensate in a PT-symmetric double well. 
arXiv:1306.3871 (2013). 

\bibitem{Ma}
Ma, Y. \& Edelman, A. 
Nongeneric eigenvalue perturbations of Jordan blocks. 
{\em Linear Algebra and its Applications} \textbf{273}, 45-63 (1998)

\bibitem{BN} 
Brody,~D. \& Rivier,~N. 
Geometrical aspects of statistical mechanics. 
{\em Physical Review} E\textbf{51}, 1006-1011 (1995). 



\end{thebibliography}
%----------------------------------------------------------

\end{document}